# Gate-tunable electronic properties of epitaxial Bi (111) films using a printable hexagonal boron nitride ionogel

Jagannath Jena[1], Heather E. Kurtz[2], Siddhesh Ambhire[3], Justin S. Wood[1], Fateme Mahdikhany[2], Junyi Yang[1], Eugene Ark[1], Vinod K. Sangwan[2], J. Samuel Jiang[1], Steven S.-L. Zhang[3], Mark C. Hersam[2,4,5], and Anand Bhattacharya[1,6]

[1]Material Science Division, Argonne National Laboratory, Lemont, IL 60439, USA

[2]Department of Materials Science and Engineering, Northwestern University, Evanston, IL 60208, USA

[3]Department of Physics, Case Western Reserve University, Cleveland, OH 44106, USA

[4]Department of Chemistry, Northwestern University, Evanston, IL 60208, USA

[5]Department of Electrical and Computer Engineering, Northwestern University, Evanston, IL 60208, USA

[6]Department of Physics and Astronomy, University of California, Riverside, CA 92521, USA

Achieving effective electrostatic control of carrier transport in semimetals remains challenging due to strong screening and multiband effects. We report efficient low-voltage top-gated control of electronic transport in epitaxial Bi (111) thin films grown on GaAs (111) substrates using a printable hexagonal boron nitride ionogel. Magnetotransport measurements reveal pronounced nonlinear Hall conductivities arising from multiband electron and hole contributions. Remarkably, the application of a small gate voltage (±0.4 V) leads to a systematic evolution of the low-field Hall conductivity slope and the electron-hole compensation point. The response to gate voltage depends upon thickness and temperature. The observed behavior cannot be explained by a conventional Fermi level shift with rigid bands and instead indicates a non-rigid band response associated with multiband effects and a gate-tunable Rashba spin–orbit coupling. Our results establish printable ionogel gating as a powerful approach to tune multiband transport in topological semimetals.

**Introduction**

Achieving efficient electrostatic control of carrier transport in semimetals has remained a longstanding challenge because of their relatively large intrinsic carrier density, short screening length, and the simultaneous presence of multiple carrier channels. In conventional metals, external electric fields are rapidly screened and therefore produce negligible modulation of bulk transport. Although semimetals possess carrier densities that are orders of magnitude lower than ordinary metals, the coexistence of electron and hole carriers often leads to a complex transport response, making controlled electrical tuning of properties challenging to interpret. Developing a reliable, high-capacitance, low-voltage gating strategy to manipulate such multiband transport is therefore of both fundamental and technological importance for gate-tunable quantum electronic devices[1-3].

Among elemental semimetals, bismuth represents an especially attractive platform owing to its exceptionally low carrier density, long Fermi wavelength, and strong spin–orbit coupling.[4-10] [11]In thin-film form, quantum confinement progressively gaps out the bulk semi-metallic states[9, 12] and enhances the contribution from highly spin-split two-dimensional surface states[5, 13-16] that remain metallic down to the lowest temperatures. The combination of low carrier density and spin-split Fermi pockets in the surface states of Bi thin films makes these electronic states potentially tunable using electrostatic gating and magnetic fields.

Indeed, recent breakthroughs have demonstrated electrostatic tunability of quantum transport in ultrathin Bi-based heterostructures.[7, 8] In particular, back-gated Bi/hBN devices have been shown to host a high-mobility two-dimensional electron system in which an applied gate-voltage ($V_g$) modifies the Hall conductivity and drives carrier redistribution under strong magnetic fields.[7] While these studies establish the feasibility of electrical tuning in confined bismuth systems, they are primarily based on back-gate geometries at large $V_g$ values (100 V), using micron-scale hBN and Bi flakes, ultimately focusing on the extreme quantum regime of van der Waals heterostructures. By contrast, top-gated low-voltage electrical control of multiband Hall transport in epitaxial Bi (111) thin films using high-capacitance gate dielectrics has remained unexplored.

From a device perspective, top gating offers significantly stronger and more localized electrostatic coupling compared to conventional back gating. We accomplish this goal using a patterned electric-double-layer in a solid-state media. Such a geometry is expected to provide efficient low-voltage modulation of the surface-dominated transport channels in Bi (111), enabling direct access to gate-controlled properties in experimentally practical Hall-bar devices. Two key unresolved questions in this direction are: Can this approach tune the relative dominance of spin-polarized electron and hole carriers, and under what thickness conditions can one carrier type be made dominant?

In this work, we demonstrate efficient low-voltage top-gated control of multiband transport in high-quality epitaxial Bi (111) thin films grown on GaAs (111) substrates using a printable, solid-state hexagonal boron nitride (hBN) ionogel gate dielectric. We show that a small $V_g$ in the range of $\pm 0.4$V can induce a pronounced and systematic evolution of the Hall conductivity, including a gate-dependent change in the low-field Hall slope and a strong shift of the compensation field $B^*$, where the electron and hole Hall contributions cancel. While thicker films exhibit gate-tunable coexistence of both carrier types, thinner films can be driven into a robust hole-dominated transport regime under positive gate voltage bias. These observations reveal an unconventional gating response beyond a simple rigid-band shift picture and establish a practical route for electrically controlling electronic structure of epitaxial bismuth thin films.

**Results and discussion**

High-quality thin films of Bi (111) were grown on GaAs (111) substrates using molecular beam epitaxy (MBE) under ultra-high vacuum conditions (base pressure ≈ $1.3 \times 10^{-10}$ Torr). The detailed growth process is described elsewhere.[17] X-ray diffraction (XRD) measurements reveal distinct peaks at 22.446°, 45.752°, and 71.319° in the 2θ–ω scan, corresponding to the Bi (111), Bi (222), and Bi (333) planes, confirming the epitaxial nature of the film (Fig. 1a). The peaks associated with the GaAs (111) substrate are marked in red. X-ray reflectivity (XRR) exhibits well-defined oscillations (Fig. 1b), while atomic force microscopy (AFM) further confirms a smooth surface morphology of the Bi (111) film (Fig. 1c). To study electronic transport, standard lithography was used to fabricate Hall bar devices. The bar width ($W$) is ≈ 200 μm, with the overall channel length and the spacing between longitudinal voltage probes ($L$) measuring ≈ 2 mm and ≈ 700 μm, respectively. Electrical contacts were formed by sputtering a wetting layer of Pt (10 nm) and then Au (70 nm).

For field-effect modulation, a large area gate electrode pad was fabricated separately on the substrate and electrically isolated from the Hall bar (Fig. 2a). This pad was used to apply $V_g$ via a printed hBN ionogel, which covered both the Hall bar and a portion of the gate electrode. Details of the hBN ionogel preparation and properties have been reported previously.[18, 19] Briefly, exfoliated hBN nanoplatelets are combined with the ionic liquid 1-ethyl-3-methylimidazolium bis(trifluoromethylsulfonyl)imide (EMIM-TFSI) in a diluent solvent of ethyl lactate to form a printable hBN ionogel ink. The hBN ionogel was deposited via aerosol-jet printing onto the Hall bar devices over an area of approximately 0.5 mm × 1 mm covering both the Hall bar and the side gate electrode. During printing, the platen stage was held at 60 °C to promote evaporation of the diluent solvent, resulting in a solid, ionogel electrolyte for precise and localized gating. This top-gated architecture provides strong localized electric-field coupling to the Bi surface states and enables efficient modulation at low $V_g$. Electronic transport measurements were carried out in a Quantum Design Physical Properties Measurement System (PPMS), with $V_g$ applied using a Keithley 2400 source meter.

As a reference, transport measurements were first performed without any applied $V_g$. Subsequently, positive and negative $V_g$ were applied to investigate the gating response of the Bi (111) film. The applied $V_g$ was limited to $\pm 0.5$ V, within which reproducible modulation of the conductivity was observed. Beyond this range, no further evolution of the conductivity curves was detected, presumably due to saturation of the electric double layer in the printed hBN ionogel device.

Figure 2b shows the Hall conductivity $\sigma_{xy}$ of a 30 nm Bi (111) film, obtained both without $V_g$ (black curve) and under applied $V_g$ (blue and orange curves). The conductivity was calculated from the measured longitudinal resistance $R_{xx}$ and transverse resistance $R_{xy}$. The corresponding resistivities were obtained using $\rho_{xx} = R_{xx} \cdot (W.t/L)$ and $\rho_{xy} = R_{xy} \cdot t$, where $W$, $L$, and $t$ are the width, length, and thickness of the Hall bar, respectively. The conductivity tensor components were then calculated by inverting the resistivity tensor using the following expressions:

$$\sigma_{xx} = \rho_{xx} \,/\, (\rho_{xx}^2 + \rho_{xy}^2)$$

$$\sigma_{xy} = -\rho_{xy} \,/\, (\rho_{xx}^2 + \rho_{xy}^2)$$

As shown in Fig. 2b, the Hall conductivity at 5 K exhibits pronounced nonlinearity with magnetic field: for positive values of field, it initially decreases, reaches a minimum at low fields, and subsequently increases, crossing zero and becoming positive at higher fields. This behavior reflects the presence of multiple electron/hole pockets contributing to transport, consistent with our previous experimental and theoretical studies[17]. The linear behavior in the low-field region, characterized by a negative slope, is attributed to high-mobility electron pockets located near the *M*-points of the Bi (111) Fermi surface[17].

With increasing magnetic field, the conductivity crosses zero at a characteristic field $B^*$, referred to as the compensation field, where the positive contribution from hole carriers exactly balances the negative contribution from electron carriers, resulting in zero net Hall conductivity. At higher fields, the conductivity becomes positive and is dominated by hole carriers.

In the absence of $V_g$, the compensation field is approximately $B^* \approx 3.03$ T. Upon application of $V_g$, $B^*$ shifts significantly depending on the polarity. In particular, $B^*$ decreases to $\approx$ 1.43 T for $+0.4$V and increases to $\approx$ 4T for $-0.4$V. In addition, the magnitude of the low-field negative slope changes systematically with $V_g$. For $+0.4$V, both $B^*$ and the magnitude of the negative low-field slope are reduced, while the high-field positive slope is enhanced compared to the ungated and $-0.4$V cases. These trends indicate that positive $V_g$ suppresses the contribution from electron pockets near the *M*-points, thereby enhancing hole-dominated transport, whereas negative $V_g$ strengthens the electron contribution. Overall, hBN ionogel gating effectively tunes the relative contributions of electrons and holes, driving the system across regimes of carrier dominance.

We note that the evolution of the Hall response cannot be explained solely by changes in carrier density arising from a rigid shift of the chemical potential within an otherwise unchanged band structure. Such a simplified electrostatic picture would predict that positive $V_g$ leads to increased electron accumulation and strengthens the surface electron contribution to the Hall resistance. Experimentally, however, the opposite trend is observed in Bi (111). The positive $V_g$ weakens the electron-dominated Hall response, shifting the compensation field $B^*$ to a lower magnetic field, whereas the opposite effect is observed when a negative $V_g$ is applied. These observations cannot be explained by a shift in chemical potential in the rigid-band approximation, pointing rather to gate induced changes in the underlying electronic band structure.

To further elucidate the role of electron and hole pockets in gating, we investigated thinner Bi films. Angle-resolved photoemission spectroscopy (ARPES) measurements on ultrathin Bi films suggest that the size of the *M*-point electron pockets is strongly reduced due to quantum confinement effects[20, 21], while the hole pockets remain comparatively robust. In our thinnest films with thickness ≈5 nm[17], we find that $\sigma_{xy}(B)$ is indeed dominated by hole carriers over the entire magnetic field range, while at thicknesses above 16 nm, $\sigma_{xy}(B)$ at low fields begins to get dominated by high-mobility electrons, whose Hall conductance increases progressively with thickness. Motivated by this, we chose to gate a 16 nm Bi (111) film.

In the absence of $V_g$, $\sigma_{xy}(B)$ for the 16 nm film exhibits a smaller low-field slope (Fig. 3a, black curve) and a reduced compensation field $B^*(\approx 2.23\ \mathrm{T})$ compared to the 30-nm-thick film, indicating a relatively diminished contribution from high-mobility electron pockets. Upon applying $V_g$, the transport response evolves systematically (Fig. 3a, blue and orange curves). Under negative $V_g$, the magnitude of the negative low-field slope increases, signifying an enhanced contribution from electron carriers. In contrast, under a positive $V_g$ of $+0.4$V, the low-field negative slope is completely suppressed, and a positive slope is observed across the entire measured field range up to 7.5 T at 5 K, indicating that transport is dominated by hole carriers throughout the field range, with no observable compensation point.

Additionally, we performed two-band conductivity fitting to extract carrier mobilities and carrier densities at different values of $V_g$ with respect to the zero-gate-voltage condition. The Hall conductivity can be expressed as $\sigma_{xy}(B) = eB\sum_i \frac{sen_i\mu_i^2}{1+(\mu_i B)^2}$, where $s = \pm$ for electrons and holes, respectively. It is important to note that, due to the inherent parameter degeneracy in the two-band model, independent extraction of all four parameters ($n_e$, $n_h$, $\mu_e$, $\mu_h$) is not unique[22-24]. Therefore, physically motivated constraints are required. In our analysis, we utilize the high-field regime ($\mu B \gg 1$), where the Hall conductivity exhibits an approximately linear dependence on magnetic field with a positive slope, indicating a dominant hole contribution. In this limit, the expression simplifies, allowing us to extract the quantity $en_h\mu_h^2$ from linear fitting of the high-field $\sigma_{xy}(B)$, although $n_h$ and $\mu_h$ cannot be independently determined. In contrast, the low-field regime is dominated by high-mobility electron carriers, as evidenced by the negative slope of $\sigma_{xy}$. In bismuth, this behavior is primarily associated with electron pockets located at the *M*-points,[17] which contribute significantly due to their high mobility. By analyzing the evolution of the low-field slope with $V_g$, we find that $en_e\mu_e^2$ for electrons is highest under negative gate bias, corresponding to the steepest negative slope, and decreases progressively for zero and positive values of $V_g$. The extracted electron mobilities for both 30 nm and 16 nm Bi films are shown in Fig. 4b whereas Fig. 4c displays the corresponding carrier densities. These results are opposite to what would be expected from electrostatic gating that merely shifts the Fermi level, where a positive $V_g$ would enhance contributions from electrons and a negative $V_g$ would enhance contributions from holes.

To gain microscopic insight into the unconventional gate dependence of the Hall response, we performed tight-binding calculations for Bi (111) thin films using a multiorbital $sp^3$ Hamiltonian with strong atomic spin–orbit coupling. The calculations reveal that orbital hybridization between the $s$, $p_x$, $p_y$, and $p_z$ orbitals become allowed by the broken inversion symmetry at the Bi (111) surface and hybridization plays a crucial role in shaping the Fermi contour. As shown in Figs. 3(b) and 3(c), inclusion of this hybridization leads to the emergence of six-hole pockets together with a substantial reduction of the electron pockets, resulting in a pronounced reconstruction of the multiband Fermi contour. Thus, a surface electric field that breaks inversion symmetry may be used to tune the balance between the electron and hole pockets.

Therefore, a plausible interpretation of our data is that the applied $V_g$ modifies the effective Rashba-type surface interaction. Within the tight-binding model, stronger inversion-asymmetry-induced orbital hybridization promotes the emergence of hole pockets, whereas weaker hybridization favors a more electron-dominated Fermi contour. Combined with the experimentally observed suppression of the electron-dominated Hall response under positive gate bias and its enhancement under negative gate bias, these results suggest that positive and negative $V_g$ affect the effective Rashba interaction in opposite directions. Establishing the absolute sign of this effect, however, is nontrivial in Bi (111) because of its complex multiorbital surface electronic structure[25, 26]. In the present tight-binding model, the effective Rashba interaction is represented indirectly through phenomenological surface-hybridization parameters, making it difficult to uniquely determine whether a given gate polarity enhances or suppresses the effective Rashba interaction.

Application of a positive $V_g$ presumably reinforces the intrinsic surface electric field and thereby strengthens the Rashba splitting. Under this condition, the tight-binding calculations show a substantial shrinkage of the electron pockets at the *M*-points together with the emergence and strengthening of hole pockets along the *Γ-M* directions (Fig. 3c). This gate-induced reconstruction of the Rashba-derived Fermi surface increases the carrier density of hole-type carriers, leading to the suppression of the negative Hall slope (e.g., Fig. 2c of 30 nm Bi film (blue curve)) and, in thinner films (Fig. 3a of 16 nm Bi film (blue curve)), the complete disappearance of the compensation field. The close correspondence between the experimentally observed polarity-dependent Hall conductivity and the theoretically predicted Rashba-field-controlled evolution of the Fermi pockets strongly indicates that the $V_g$ does not simply cause a rigid Fermi-level shift, but modifies the net surface electric field and hence the multiband Rashba electronic structure of Bi (111).

To gain further insight, we performed temperature-dependent Hall conductivity measurements (Fig. 3d-f). For the positive gate bias (+0.4 V), the Hall conductivity maintains a positive slope up to 40 K without any observable compensation field, confirming robust hole-dominated transport. In contrast, for zero and negative values of $V_g$, the compensation field is evident irrespective of temperatures, and the magnitude of the negative low-field slope increases with increasing temperature, reflecting the contribution of high-mobility electron carriers (Fig. 4d, 4e). Nominally, the high mobility electrons are expected to become gapped at low thicknesses. However, applying a negative gate voltage, and increasing temperature can both give rise to high mobility electrons, implying that temperature and $V_g$ can both be used to induce electron like carriers across the gap. These results provide further insight about how $V_g$ can effectively tune the system between mixed-carrier and hole-dominated transport regimes depending on the polarity of the applied voltage.

Finally, it is worth emphasizing that this gating response differs fundamentally from the conventional rigid-band electrostatic picture typically observed in semimetals such as graphene or $Cd_3As_2$, where positive $V_g$ usually promotes electron accumulation and negative $V_g$ enhances hole transport.[27, 28] In contrast, our Bi thin films exhibit an opposite trend, where a positive $V_g$ leads to hole-dominated transport, and a negative $V_g$ enhances the contribution from electron-like carriers. This unconventional gating response indicates that the effect of the electric field in our system cannot be described by a simple rigid-band picture. Instead, it suggests that gating modifies the relative contributions of electron and hole pockets, likely through changes in band structure and carrier redistribution associated with strong spin–orbit coupling and multiband effects in bismuth.

## Conclusions

In summary, we have demonstrated gate-controlled modulation of multiband transport in epitaxial Bi(111) thin films using a printable hBN ionogel gating platform. The Hall conductivity exhibits strong nonlinearity due to competing electron and hole contributions, and both the low-field slope and the compensation field ($B^*$) can be systematically tuned with $V_g$. In 30 nm Bi (111) films, gating modifies the relative balance between carriers, while in thinner (16 nm) films, a positive $V_g$ suppresses the electron contribution entirely, leading to robust hole-dominated transport across the full magnetic field range. The observed gate response deviates from the conventional rigid-band picture and instead reflects a redistribution of carriers and possible band-structure modification driven by strong spin–orbit coupling and multiband effects in bismuth. These results establish a pathway for controlling carrier dominance in elemental semimetals and highlight the effectiveness of low-voltage gating of solid-state ionogel for engineering electronic transport in quantum materials.

## Acknowledgements

This research was supported by the DOE ASCR Microelectronics Science Research Center Projects BIA, which is supported by the U.S. Department of Energy, Office of Science, under contract number DE-AC02-06CH11357. AB acknowledges support from the University of California at Riverside during the preparation of this manuscript.

## Figures

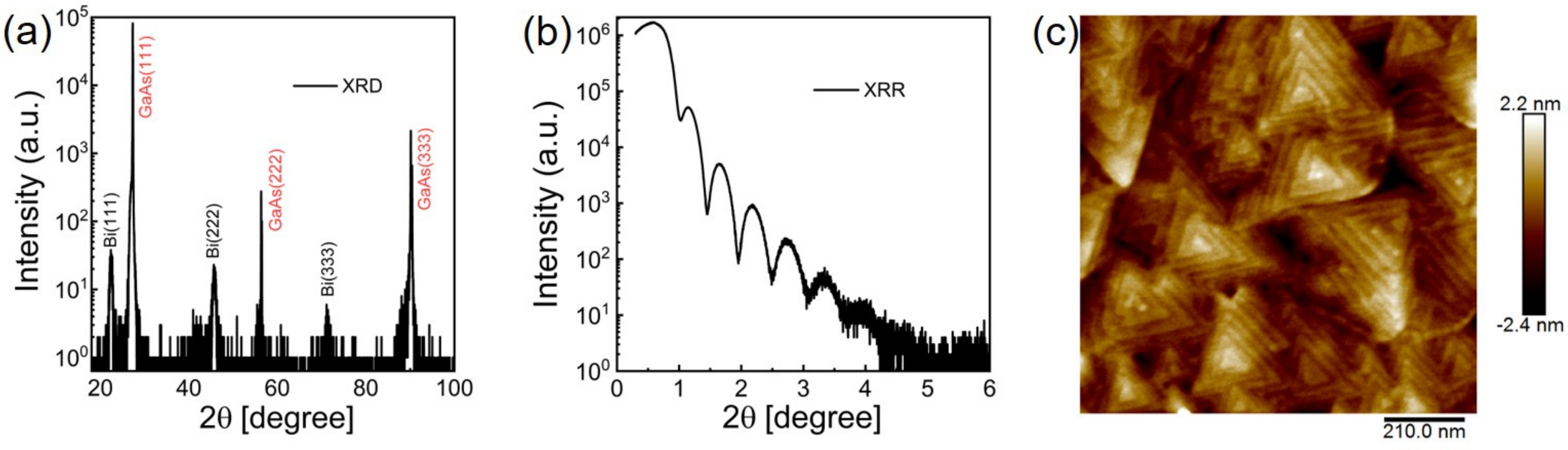

**Figure 1:** (a) X-ray diffraction (XRD) pattern of a 16 nm Bi (111) thin film. The diffraction peaks corresponding to the Bi(111) planes are labeled in black, while the substrate peaks are indicated in red. (b) X-ray reflectivity (XRR) profile of the same Bi film. (c) Atomic force microscopy (AFM) image showing the surface morphology of the film, with a measured root-mean-square (RMS) roughness of 0.520 nm, indicating a smooth and uniform surface.

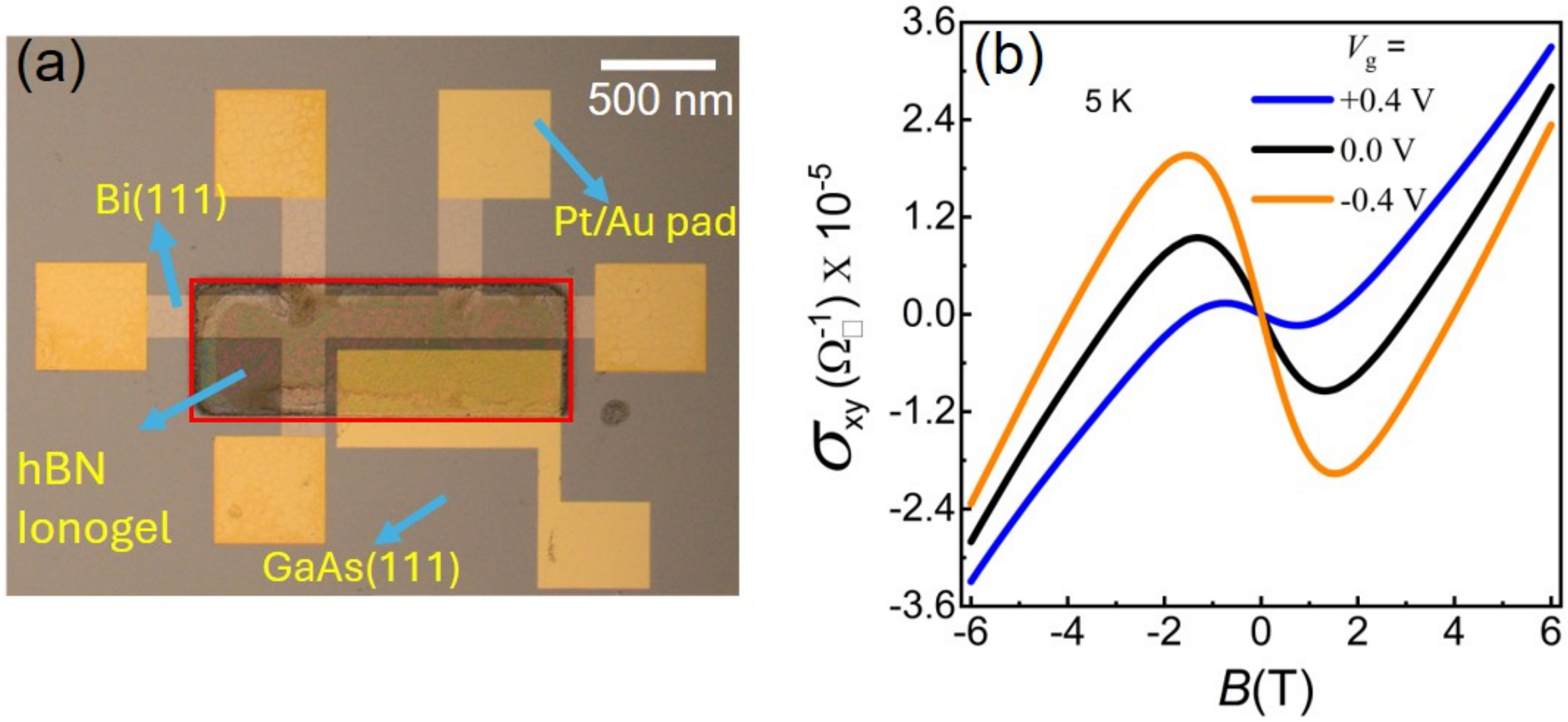

**Figure 2:** (a) Optical image of the ionogel printed device. The Au/Pt contact pads, Bi thin-film Hall bar, and the gate electrode are indicated by arrows. The region covered by the printed hBN ionogel is highlighted by the red rectangle. (b) Hall conductivity measured without a gate voltage bias ($V_g$) application (black line), illustrating the baseline transport behavior of the device.

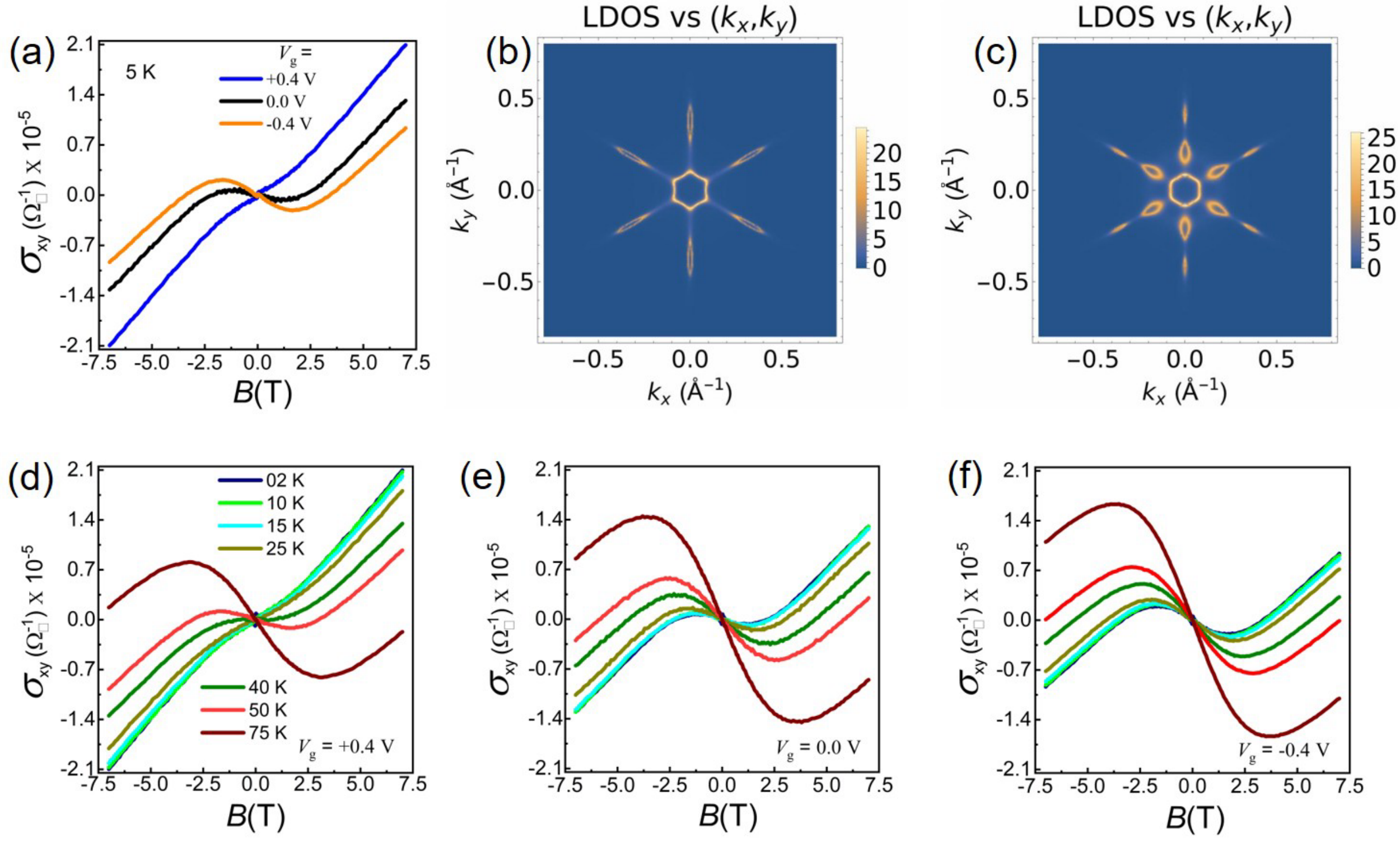

**Figure 3:** (a) Comparison of Hall conductivity for a 16 nm Bi film under various $V_g$ at 5 K. The orange, black, and purple curves correspond to $V_g$ of +0.4 V, 0 V, and −0.4 V, respectively, illustrating clear modulation of electronic transport behavior by gate control. (b) Calculated Fermi surface contour in the absence of inter-orbital hybridization between the $s$, $p_x$, $p_y$, and $p_z$ orbitals for a 6 bilayer (1 bilayer = 0.39Aº) Bi, showing a hexagonal electron pocket centered at *Γ*-point and six large needle-shaped electron pockets near *M*-points. (c) Fermi contour highlighting the emergence of six-hole pockets along the *Γ*–*M* direction when inter-orbital hybridization between the $s$, $p_x$, $p_y$, and $p_z$ orbitals is taken into account, allowed by the broken inversion symmetry near the Bi (111) surface. These hole pockets lie between the hexagonal electron pocket centered at the *Γ*-point and the electron pockets near the *M*-points. The latter are reduced in size (d–f). Temperature-dependent Hall conductivity (2–75 K) under different gate voltage biases: +0.4 V (c), 0 V (d), and −0.4 V (e), showing the evolution of transport features.

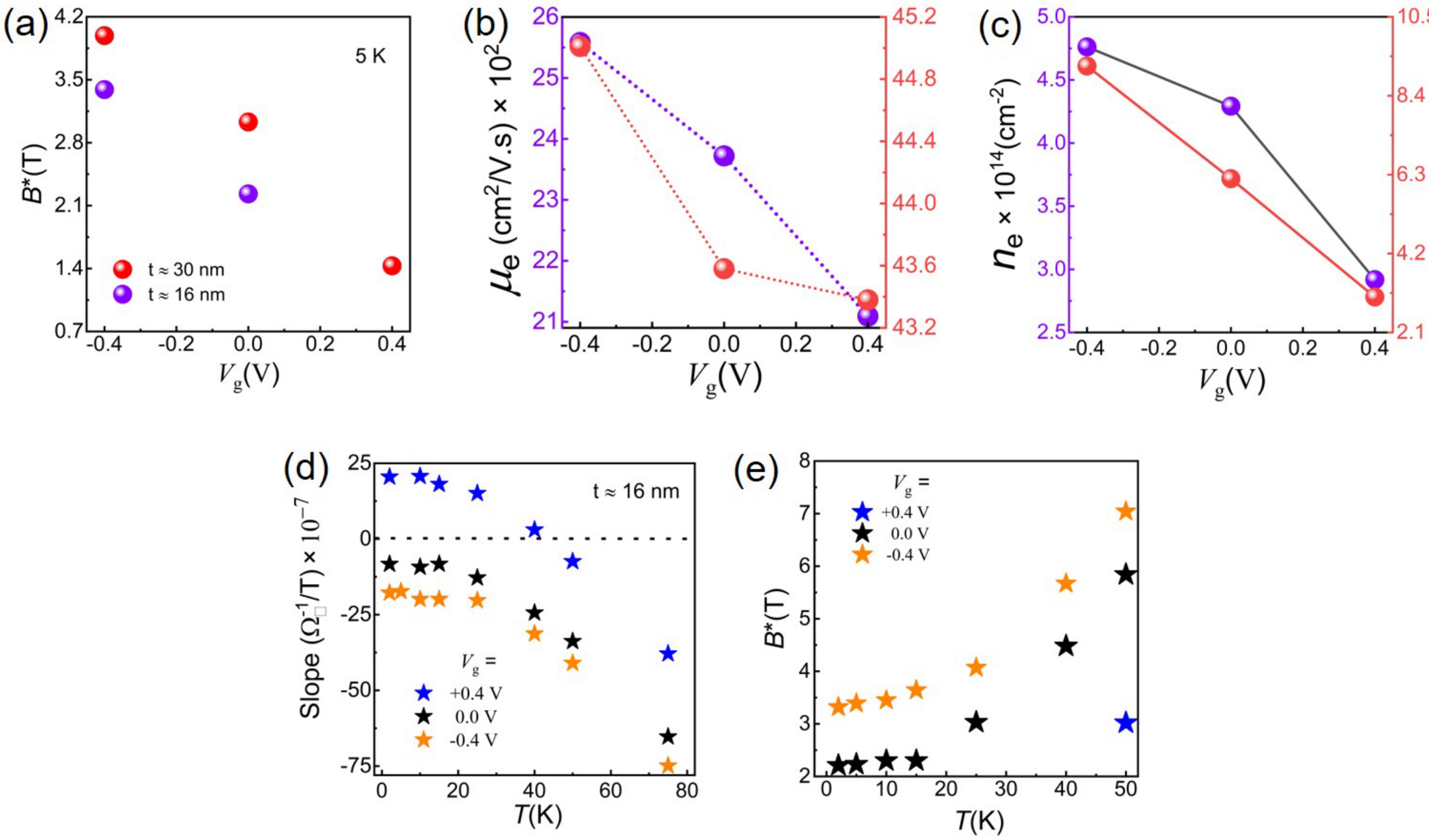

**Figure 4:** (a) Zero-crossing conductivity field as a function of $V_g$ for 30 nm (red) and 16 nm Bi (purple) films at 5 K. (b, c) Electron mobility and carrier density obtained from two-band model fitting for the conductivity of 16 nm and 32 nm Bi films as functions of $V_g$. (d) Low-field slopes extracted from linear fitting of the conductivity curves of 16 nm Bi film close to zero magnetic field. (e) Zero-crossing conductivity field as a function of temperature in a 16 nm Bi film. The blue, black, and orange stars are for $V_g$ of +0.4 V, 0.0 V, and -0.4 V, respectively.

## References

(1) Zhai, E.; Liang, T.; Liu, R.; Cai, M.; Li, R.; Shao, Q.; Su, C.; Lin, Y. C. The rise of semi-metal electronics. *Nature Reviews Electrical Engineering* **2024**, *1* (8), 497–515. DOI: https://doi.org/10.1038/s44287-024-00068-z.

(2) Hu, J.; Xu, S.-Y.; Ni, N.; Mao, Z. Transport of topological semimetals. *Annual Review of Materials Research* **2019**, *49* (1), 207–252. DOI: https://doi.org/10.1146/annurev-matsci-070218-010023.

(3) Burkov, A. Topological semimetals. *Nature materials* **2016**, *15* (11), 1145–1148. DOI: https://doi.org/10.1038/nmat4788.

(4) Edelman, V. S. Electrons in Bismuth. *Adv Phys* **1976**, *25* (6), 555–613. DOI: https://doi.org/10.1080/00018737600101452.

(5) Hofmann, P. The surfaces of bismuth: Structural and electronic properties. *Prog Surf Sci* **2006**, *81* (5), 191–245. DOI: https://doi.org/10.1016/j.progsurf.2006.03.001.

(6) Makushko, P.; Kovalev, S.; Zabila, Y.; Ilyakov, I.; Ponomaryov, A.; Arshad, A.; Prajapati, G. L.; De Oliveira, T. V.; Deinert, J.-C.; Chekhonin, P. A tunable room-temperature nonlinear Hall effect in elemental bismuth thin films. *Nature Electronics* **2024**, *7* (3), 207–215. DOI: https://doi.org/10.1038/s41928-024-01118-y.

(7) Yue, D.; Wang, H.; Huang, G.; Jiang, Y.; Huang, Z.; Zheng, P.; Song, Y.; Guo, S.; Tian, N.; Luo, M. Quantum Transport in Bismuth Two-Dimensional Electron System. *Physical Review X* **2025**, *15* (4), 041047. DOI: DOI: https://doi.org/10.1103/hptx-pw9s.

(8) Chen, L.; Wu, A. X.; Tulu, N.; Wang, J.; Juanson, A.; Watanabe, K.; Taniguchi, T.; Pettes, M. T.; Campbell, M. A.; Xu, M. Exceptional electronic transport and quantum oscillations in thin bismuth crystals grown inside van der Waals materials. *Nature Materials* **2024**, *23* (6), 741–746. DOI: https://doi.org/10.1038/s41563-024-01894-0.

(9) Ito, S.; Feng, B.-j.; Arita, M.; Takayama, A.; Liu, R.-Y.; Someya, T.; Chen, W.-C.; Iimori, T.; Namatame, H.; Taniguchi, M. Proving nontrivial topology of pure bismuth by quantum confinement. *Phys Rev Lett* **2016**, *117* (23), 236402. DOI: https://doi.org/10.1103/PhysRevLett.117.236402.

(10) Seo, J.; Guo, C. M.; Putzke, C.; Huang, X.; Goodge, B. H.; Wong, Y. C.; Fischer, M. H.; Neupert, T.; Moll, P. J. Transport evidence for chiral surface states from three-dimensional Landau bands. *Nature Physics* **2026**, 1–7. DOI: https://doi.org/10.1038/s41567-025-03146-7.

(11) Yánez-Parreño, W. J.; Vera, A.; Santhosh, S.; Dong, C.; Kotsakidis, J. C.; Ou, Y.; Islam, S.; Friedman, A. L.; Wetherington, M.; Robinson, J. Charge-to-spin conversion in atomically thin bismuth. *Physical Review Applied* **2025**, *23* (6), L061001. DOI: https://doi.org/10.1103/fwrg-3dr2.

(12) Hoffman, C.; Meyer, J.; Bartoli, F.; Di Venere, A.; Yi, X.; Hou, C.; Wang, H.; Ketterson, J.; Wong, G. Semimetal-to-semiconductor transition in bismuth thin films. *Physical Review B* **1993**, *48* (15), 11431. DOI: https://doi.org/10.1103/PhysRevB.48.11431.

(13) Ast, C. R.; Höchst, H. Fermi surface of Bi(111) measured by photoemission spectroscopy -: art. no. 177602. *Phys Rev Lett* **2001**, *87* (17). DOI: https://doi.org/10.1103/PhysRevLett.87.177602.

(14) Hirahara, T.; Nagao, T.; Matsuda, I.; Bihlmayer, G.; Chulkov, E. V.; Koroteev, Y. M.; Echenique, P. M.; Saito, M.; Hasegawa, S. Role of spin-orbit coupling and hybridization effects in the electronic structure of ultrathin Bi films. *Phys Rev Lett* **2006**, *97* (14). DOI: https://doi.org/10.1103/PhysRevLett.97.146803.

(15) Ito, S.; Arita, M.; Haruyama, J.; Feng, B.; Chen, W. C.; Namatame, H.; Taniguchi, M.; Cheng, C. M.; Bian, G.; Tang, S. J.; et al. Surface-state Coulomb repulsion accelerates a metal-insulator transition in topological semimetal nanofilms. *Sci Adv* **2020**, *6* (12). DOI: https://doi.org/10.1126/sciadv.aaz5015.

(16) Takayama, A.; Sato, T.; Souma, S.; Oguchi, T.; Takahashi, T. Tunable Spin Polarization in Bismuth Ultrathin Film on Si(111). *Nano Lett* **2012**, *12* (4), 1776–1779. DOI: https://doi.org/10.1021/nl2035018.

(17) Jena, J.; Ark, E. D.; Ambhire, S.; Smith, M. D.; Wood, J. S.; Yang, J.; Pearson, J.; Arava, H.; Rosenmann, D.; Welp, U. Synthesis and transport properties of epitaxial Bi (111) films on GaAs (111) substrates. *APL Materials* **2026**, *14* (2). DOI: https://doi.org/10.1063/5.0304277.

(18) Hyun, W. J.; Chaney, L. E.; Downing, J. R.; de Moraes, A. C.; Hersam, M. C. Printable hexagonal boron nitride ionogels. *Faraday Discussions* **2021**, *227*, 92–104. DOI: https://doi.org/10.1039/c9fd00113a.

(19) Thomas, C. M.; Hyun, W. J.; Huang, H. C.; Zeng, D.; Hersam, M. C. Blade-coatable hexagonal boron nitride ionogel electrolytes for scalable production of lithium metal batteries. *ACS Energy Letters* **2022**, *7* (4), 1558–1565. DOI: https://doi.org/10.1021/acsenergylett.2c00535.

(20) Bian, G.; Wang, X.; Kowalczyk, P. J.; Maerkl, T.; Brown, S. A.; Chiang, T.-C. Survey of electronic structure of Bi and Sb thin films by first-principles calculations and photoemission measurements. *Journal of Physics and Chemistry of Solids* **2019**, *128*, 109–117. DOI: https://doi.org/10.1016/j.jpcs.2017.07.027.

(21) Miao, L.; Yao, M.-Y.; Ming, W.; Zhu, F.; Han, C. Q.; Wang, Z. F.; Guan, D. D.; Gao, C. L.; Liu, C.; Liu, F.; et al. Evolution of the electronic structure in ultrathin Bi(111) films. *Physical Review B* **2015**, *91* (20), 205414. DOI: 10.1103/PhysRevB.91.205414.

(22) Ashcroft, N. W.; Mermin, N. D. Solid state physics. holt, rinehart and winston, new york London: 1976.

(23) Kim, J. S.; Seiler, D. G.; Tseng, W. Multicarrier characterization method for extracting mobilities and carrier densities of semiconductors from variable magnetic field measurements. *Journal of applied physics* **1993**, *73* (12), 8324–8335. DOI: https://doi.org/10.1063/1.353424.

(24) Dill, J. E.; Chang, C. F.; Jena, D.; Xing, H. G. Two-carrier model-fitting of Hall effect in semiconductors with dual-band occupation: A case study in GaN two-dimensional hole gas. *Journal of Applied Physics* **2025**, *137* (2). DOI: https://doi.org/10.1063/5.0248998.

(25) Ast, C. R.; Gierz, I. sp-band tight-binding model for the Bychkov-Rashba effect in a two-dimensional electron system including nearest-neighbor contributions from an electric field. *Physical Review B—Condensed Matter and Materials Physics* **2012**, *86* (8), 085105. DOI: https://doi.org/10.1103/PhysRevB.86.085105.

(26) Tokatly, I.; Krasovskii, E.; Vignale, G. Current-induced spin polarization at the surface of metallic films: A theorem and an ab initio calculation. *Physical Review B* **2015**, *91* (3), 035403. DOI: https://doi.org/10.1103/PhysRevB.91.035403.

(27) Liu, Y.; Zhang, C.; Yuan, X.; Lei, T.; Wang, C.; Di Sante, D.; Narayan, A.; He, L.; Picozzi, S.; Sanvito, S. Gate-tunable quantum oscillations in ambipolar $Cd_3As_2$ thin films. *NPG Asia Materials* **2015**, *7* (10), e221–e221. DOI: https://doi.org/10.1038/am.2015.110.

(28) Katsnelson, M. I.; Novoselov, K. S.; Geim, A. K. Chiral tunnelling and the Klein paradox in graphene. *Nature physics* **2006**, *2* (9), 620–625. DOI: https://doi.org/10.1038/nphys384.